\documentclass[twocolumn]{aastex7}

\newcommand{\Msun}{\,{\rm M_\odot}}

\begin{document}

\title{Cosmic Outliers: Low-Spin Halos Explain the \\ Abundance, Compactness, and Redshift Evolution of the Little Red Dots}

\correspondingauthor{Fabio Pacucci} 

\author[orcid=0000-0001-9879-7780]{Fabio Pacucci}
\affiliation{Center for Astrophysics $\vert$ Harvard \& Smithsonian, 60 Garden St, Cambridge, MA 02138, USA}
\affiliation{Black Hole Initiative, Harvard University, 20 Garden St, Cambridge, MA 02138, USA}
\email[show]{fabio.pacucci@cfa.harvard.edu}

\author[0000-0003-4330-287X]{Abraham Loeb}
\affiliation{Center for Astrophysics $\vert$ Harvard \& Smithsonian, 60 Garden St, Cambridge, MA 02138, USA}
\affiliation{Black Hole Initiative, Harvard University, 20 Garden St, Cambridge, MA 02138, USA}
\email{aloeb@cfa.harvard.edu}

\begin{abstract}
The Little Red Dots (LRDs) are high-redshift galaxies uncovered by JWST, characterized by small effective radii ($R_{\rm eff} \sim 80-300$ pc), number densities that are intermediate between those of typical galaxies and quasars, and a redshift distribution peaked at $z \sim 5$.
We present a theoretical model in which the LRDs descend from dark matter halos in the extreme low-spin tail of the angular momentum distribution. Within this framework, we explain their three key observational signatures: (i) abundance, (ii) compactness, and (iii) redshift distribution.
Our model focuses on observed, not modeled, properties; it is thus independent of whether they are powered primarily by a black hole or stars.
We find that the assumption that the prototypical LRD at $z\sim5$ originates from halos in the lowest $\sim 1\%$ of the spin distribution is sufficient to reproduce both their observed number densities and physical sizes. The redshift evolution of their observability is driven by the interplay between the evolving compact disk fraction and cosmological surface brightness dimming. This effect leads to a well-defined ``LRDs Era'' at $4<z<8$, during which the LRDs are common and detectable; at $z<4$, they are bright but rare, while at $z>8$, they are common but faint.
Finally, we test the predicted redshift trend against observational data, finding excellent agreement. 
Additional observational support comes from their excess small-scale clustering and spectral signatures of extreme core densities, both of which are expected outcomes of galaxy formation in low-spin halos. These findings suggest that the LRDs are not a fundamentally distinct population but the natural manifestation of galaxies forming in the rarest, lowest angular momentum environments.
\end{abstract}

\keywords{\uat{Early universe}{435} ---\uat{Galaxies}{573} --- \uat{Cosmology}{343} --- \uat{Galaxy evolution}{594} --- \uat{Active galaxies}{17}}


\section{Introduction}
\label{sec:introduction}

The German physicist Werner Heisenberg once remarked: ``What we observe is not nature itself, but nature exposed to our method of questioning'' \citep{Heisenberg_1958}. While rooted in quantum mechanics, this principle also describes the transformation that the James Webb Space Telescope (JWST) has brought to the field of extragalactic astronomy.
In just a few years, JWST has dramatically expanded our view of the early Universe,  detecting $z > 14$ galaxies (e.g., \citealt{Carniani_2024}), and uncovering a new population of high-redshift sources that had eluded earlier telescopes. These objects, named the Little Red Dots (LRDs; \citealt{Kocevski_2023, Harikane_2023, Matthee_2023}), are compact and red. 

\begin{figure*}[ht!]
\plotone{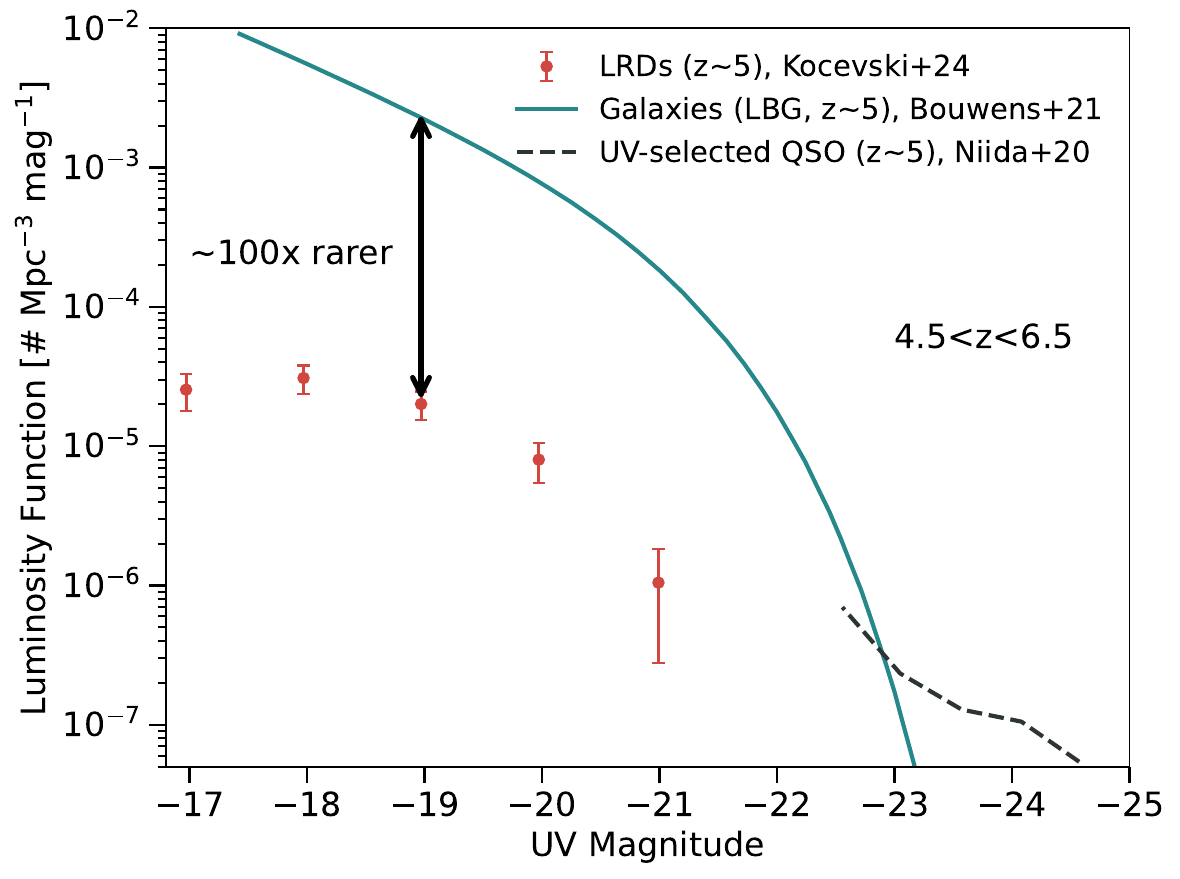}
\caption{Comparison of rest-frame UV luminosity functions for different astrophysical populations at $4.5 < z < 6.5$, spanning absolute UV magnitudes $-17 < M_{\rm UV} < -25$. Red points show the number density of LRDs from \citet{Kocevski_2024}; the dashed line indicates UV-selected quasars from \citet{Niida_2020}; the solid line shows Lyman-break galaxies (LBGs) from \citet{Bouwens_2021}. LRDs occupy an intermediate number density regime: $\sim 100$ times rarer than typical star-forming galaxies but more abundant than bright quasars.
\label{fig:n_comparison}}
\end{figure*}

The LRDs populate the early Universe with a surprising abundance: they are characterized by a number density at $z > 4$ that is intermediate between that of quasars and standard, Lyman-break galaxies. Several studies (see, e.g., \citealt{Kocevski_2023, Kokorev_2024_census}) report number densities at $4.5 < z < 6.5$ of $\sim 3\times 10^{-5} - 10^{-6} \, \mathrm{Mpc^{-3} \, mag^{-1}}$, which are $\sim 2$ orders of magnitude lower than standard galaxies, and higher than the extrapolation at fainter magnitudes of UV-selected quasars (see Fig.~\ref{fig:n_comparison}).

Detections of LRDs concentrate in the redshift range $4 < z < 8$ \citep{Kocevski_2024}, making this population visible during a cosmic period of $\sim 1$ Gyr. Recent ground-based studies suggest that LRD analogs exist at $z < 4$, but their number density appears to decline rapidly at lower redshift \citep{Ma_2025, Zhuang_2025}.

The spectral energy distribution (SED) of the LRDs typically exhibits broad Balmer emission lines, which are generally associated with the presence of an accreting supermassive black hole (SMBH) at their core \citep{Maiolino_2023_new, Greene_2023}. Remarkably, several studies have suggested that the widths of these broad lines can also be explained by the stellar velocity dispersion at the cores of these galaxies \citep{Loeb_2024RNAAS, Baggen_2024}, naturally generating velocities of $\sim 1500 \, \mathrm{km \, s^{-1}}$. In particular, \cite{Loeb_2024RNAAS} suggested that such an exceptional velocity dispersion can be caused by compact galaxies, which lack rotational support because of their low spin. This concept builds on the original idea proposed by \cite{Eisenstein_Loeb_1995} that black hole seeds originated from the collapse of overdense regions with unusually low spins.

The SED of the LRDs exhibits, in many cases, a V-shaped change in slope characterized by a red component at longer wavelengths, accompanied by a bluer component at shorter wavelengths. Remarkably, the inflection point in the SED is associated in most cases with the Balmer limit at $\lambda = 0.3645 \, \mathrm{\mu m}$ \citep{Setton_2024}.
The peculiar spectral properties of the LRDs have been thus far explained with models where most of the light comes either from a black hole or from stars (see, e.g., \citealt{Labbe_2023, Baggen_2023, Baggen_2024, Greene_2023, Kocevski_2023, Kocevski_2024, Kokorev_2023, Kokorev_2024_census, Akins_2024, Taylor_2024}).

Each interpretation faces its challenges. In the first case, the black holes seem to be overmassive \citep{Pacucci_2023_JWST, Pacucci_2024_z_evolution, Inayoshi_2024, Durodola_2024} and are undetected in X-rays, even in deep stacking analyses \citep{Ananna_2024, Yue_2024_Xray, Pacucci_Narayan_2024, Lambrides_2024, Madau_2024, Maiolino_2024_Xray}. 
In the second case, the cores reach extreme stellar densities as a direct consequence of one of the most striking features of the LRDs: their compactness. Their typical effective radii are $R_{\rm eff} \approx 150$ pc, with values varying between $80$ pc and $300$ pc \citep{Baggen_2023}.
The star-only interpretation generally requires stellar masses $ > 10^8 \Msun$ and up to $ \sim 10^{11} \Msun$, leading to immense core stellar densities \citep{Guia_2024, Baggen_2024}, up to $\sim 10^8 \Msun \, \mathrm{pc^{-3}}$, which is far above that reached at the cores of globular clusters \citep{Hopkins_2010}, and $\sim 10$ times higher than the density necessary for runaway stellar collisions to take place \citep{Ardi_2008, Fujii_2024, Vergara_2025}. It is the combination of their compactness and substantial stellar masses that renders the LRDs, in the star-only interpretation, particularly peculiar.

This Letter focuses on three key observational signatures of the LRDs: (i) their abundance, intermediate between galaxies and quasars, (ii) their compactness, and (iii) their redshift distribution, peaking at $z \sim 5$.
We demonstrate that a single physical origin can explain all three properties: the LRDs form in the \textit{low-spin tail} of the halo angular momentum distribution.

The theoretical model is described in Sec. \ref{sec:theory} and applied to the case of the LRDs in Sec. \ref{sec:results}. Section \ref{sec:disc_concl} concludes the Letter with a broader discussion on paths forward and the implications of our findings.

\section{Theoretical Framework}
\label{sec:theory}
First, we develop the theoretical model that links the three key observational signatures of the LRDs (i.e., abundance, compactness, and redshift distribution) to the spin of the progenitor dark matter halo.

\subsection{Definition of Compactness and Disk Size Model}

The LRDs are very compact galaxies, with effective radii $R_{\rm eff} < 300$ pc \citep{Baggen_2023}.
To relate compactness to halo properties, we adopt the framework developed by \citet{MMW_1998}, where the structural properties of disk galaxies are directly linked to the dimensionless spin parameter $\lambda$ of their host dark matter halo, formally defined as:
\begin{equation}
    \lambda = \frac{J_{h} |E|^{1/2}}{G M_{h}^{5/2}} \, ,
\end{equation}
where $J_{h}$ is the total angular momentum of the halo, $E$ its total energy, $M_{h}$ its mass, and $G$ is the gravitational constant \citep{BL01}. 

The scale length $R_d$ of a rotationally supported exponential disk is given by:
\begin{equation}
    R_d = \frac{1}{\sqrt{2}} \left( \frac{j_d}{m_d} \right) \lambda \, r_{200} \, ,
\end{equation}
where $r_{200}$ is the virial radius of the halo (defined as the radius enclosing a mean density that is $200$ times the critical density of the Universe at redshift $z$), $j_d = J_{\rm d}/J_{\rm h}$ is the fraction of the halo angular momentum acquired by the disk, and $m_d = M_{\rm d}/M_{\rm h}$ is the fraction of the halo mass that settles into the disk.

We further assume that baryons retain their specific angular momentum during the collapse, i.e., $j_d/m_d = 1$, which corresponds to the case where the disk forms without significant angular momentum loss (or redistribution).
This assumption is generally invalid in the low-redshift Universe when not all the baryons necessarily contribute to forming the galaxy \citep{Dutton_2007}; in this case, the angular momentum distribution does not directly translate from the halo to the galaxy. However, in the high-redshift Universe, it is more likely that most of the baryons available will form the galaxy, such that the spin distribution of the dark matter halo translates directly to that of the galaxy \citep{Romeo_2020, Romeo_2023}.
The assumption $j_d/m_d = 1$ yields a direct proportionality between $R_d$ and the spin parameter $\lambda$. 

Assuming $R_d \approx R_{\rm eff}$ (which, for an exponential disk, is correct within a factor of $\approx 1.68$, see, e.g., \citealt{NFW_1996, Elmgreen_2013}), we obtain:
\begin{equation}
    R_{\mathrm{eff}} = \frac{1}{\sqrt{2}} \lambda \, r_{200}(M_h, z) \, .
    \label{eq:size_relation}
\end{equation}

The virial radius, which is a function of total halo mass and redshift \citep{BL01}, is calculated as:
\begin{equation}
    r_{200}(M_h, z) = \left( \frac{G M_h}{100 H^2(z)} \right)^{1/3} \, ,
\end{equation}
with the Hubble parameter at redshift $z$ given by:
\begin{equation}
    H(z) = H_0 \sqrt{\Omega_m (1 + z)^3 + \Omega_\Lambda} \, .
\end{equation}
This formalism implies that at fixed halo mass, $r_{200} \propto (1 + z)^{-1}$, so that galaxy sizes shrink with increasing redshift, even for a fixed spin.

To identify which halos are capable of hosting compact LRD-like systems, we invert the size relation in Eq. (\ref{eq:size_relation}) and solve for the critical spin parameter $\lambda_{\mathrm{LRD}}$ required to produce a galaxy with a compact $R_{\mathrm{eff}}$:
\begin{equation}
    \lambda_{\mathrm{LRD}}(z) = \frac{\sqrt{2} \cdot R_{\mathrm{eff}}}{r_{200}(M_h, z)} \, .
    \label{eq:lambda_LRD}
\end{equation}

This critical spin threshold varies with redshift, becoming \textit{more permissive} (i.e., larger $\lambda_{\mathrm{LRD}}$) at higher redshift due to the shrinking virial radius. As a result, a greater fraction of halos (i.e., those with sufficiently low spin) can produce compact galaxies at early times.

\subsection{Spin Distribution and Compact Fraction}

Cosmological N-body simulations have shown that $\lambda$ follows a lognormal distribution, which is nearly independent of halo mass and redshift \citep{Warren_1992, Steinmetz_1995, Cole_Lacey_1996, MMW_1998}. We model the probability density function of the spin parameter as follows:
\begin{equation}
    p(\lambda) = \frac{1}{\lambda \sqrt{2\pi} \sigma_{\ln \lambda}} \exp \left[ -\frac{\left( \ln(\lambda/\bar{\lambda}) \right)^2}{2\sigma_{\ln \lambda}^2} \right] \, ,
    \label{eq:spin_PDF}
\end{equation}
where $\bar{\lambda}$ is the median spin parameter and $\sigma_{\ln \lambda}$ is the dispersion in its natural logarithm.
For this work, we adopt the fiducial values of $\bar{\lambda} = 0.05$ and $\sigma_{\ln \lambda} = 0.5$, used in \cite{MMW_1998}, which are consistent with high-resolution dark matter simulations.

Given a redshift-dependent spin threshold $\lambda_{\mathrm{LRD}}(z)$, the fraction of halos with spin below this threshold is calculated as the cumulative probability:
\begin{equation}
    f(\lambda < \lambda_{\mathrm{LRD}}) = \int_0^{\lambda_{\mathrm{LRD}}(z)} p(\lambda) \, d\lambda \, .
\end{equation}
This quantity represents the fraction of halos capable of producing galaxies more compact than a chosen size threshold. In our model, the redshift evolution of this compact fraction is entirely governed by the evolution of the spin threshold $\lambda_{\mathrm{LRD}}(z)$ as defined in Eq. (\ref{eq:lambda_LRD}), which accounts for the redshift-dependent virial radius.

\subsection{Surface Brightness and Observability}

Generally, compact galaxies are more likely to form in low-spin halos. However, their detectability (with, e.g., JWST) will ultimately depend on their surface brightness and how that compares with the background noise level of a given instrument. To determine whether a compact system can be observed in deep JWST imaging, we compute the apparent surface brightness as a function of redshift for galaxies with a fixed absolute magnitude (which can be related to halo mass via abundance matching, see, e.g., \citealt{Behroozi_2013, Moster_2013}) and a small effective radius.

The apparent magnitude of a galaxy with UV absolute magnitude $M_{\mathrm{UV}}$ at redshift $z$ is defined as:
\begin{equation}
    m_{\mathrm{UV}}(z) = M_{\mathrm{UV}} + 5 \log_{10} \left( \frac{D_L(z)}{10\,\mathrm{pc}} \right) \, ,
\end{equation}
where $D_L(z)$ is the luminosity distance.

The observed surface brightness $\mu$ is computed by converting the galaxy's effective radius into angular units, assuming \cite{Planck_2018} cosmology. The mean surface brightness ($\rm mag/arcsec^2$), assuming a Gaussian profile, is then given by:
\begin{equation}
    \mu(z) = m_{\mathrm{UV}}(z) + 2.5 \log_{10} \left( 2\pi R_{\mathrm{eff}}^2 \right) \, .
\end{equation}

We compare the computed surface brightness to an empirical detection limit to determine whether these sources can be detected in JWST deep imaging surveys. We adopt a $5\sigma$ detection limit for a point source of $m = 29$ mag from the CEERS survey \citep{Finkelstein_2025} and a typical NIRCam point-spread function (PSF) with radius $r = 0.07''$ \citep{Rieke_2023}. The corresponding mean surface brightness is:
\begin{equation}
    \mu_{\rm lim} \approx 25.2 \, \mathrm{mag/arcsec^2} \, .
\end{equation}

This threshold represents an absolute limit for the detectability of marginally resolved compact sources in JWST deep imaging. In this framework, sources with $\mu(z) < \mu_{\rm lim}$ are considered detectable, while those with fainter surface brightness are assumed to fall below the observational sensitivity threshold. 
Note that this is an approximate detectability threshold for unresolved or marginally resolved sources, and extended sources with lower surface brightness may be harder to detect even if $\mu < \mu_{\rm lim}$ because noise scales with area. Additionally, practical limitations (such as lack of filter coverage) can further hinder detections at high redshifts (see Sec. \ref{subsubsec:common_plus_faint}).

\section{Results}
\label{sec:results}
Next, we apply the theoretical framework developed in Sec.~\ref{sec:theory} to the case of the LRDs. We demonstrate that their origin in the low-spin population of host halos explains, \textit{simultaneously}, their (i) abundance in Sec.~\ref{subsec:abundance}, (ii) compactness in Sec.~\ref{subsec:compactness}, and (iii) redshift distribution in Sec.~\ref{subsec:z_evol}, where we also test our models with observational data.

\subsection{Abundance of the LRDs}
\label{subsec:abundance}

\begin{figure*}[ht!]
\plotone{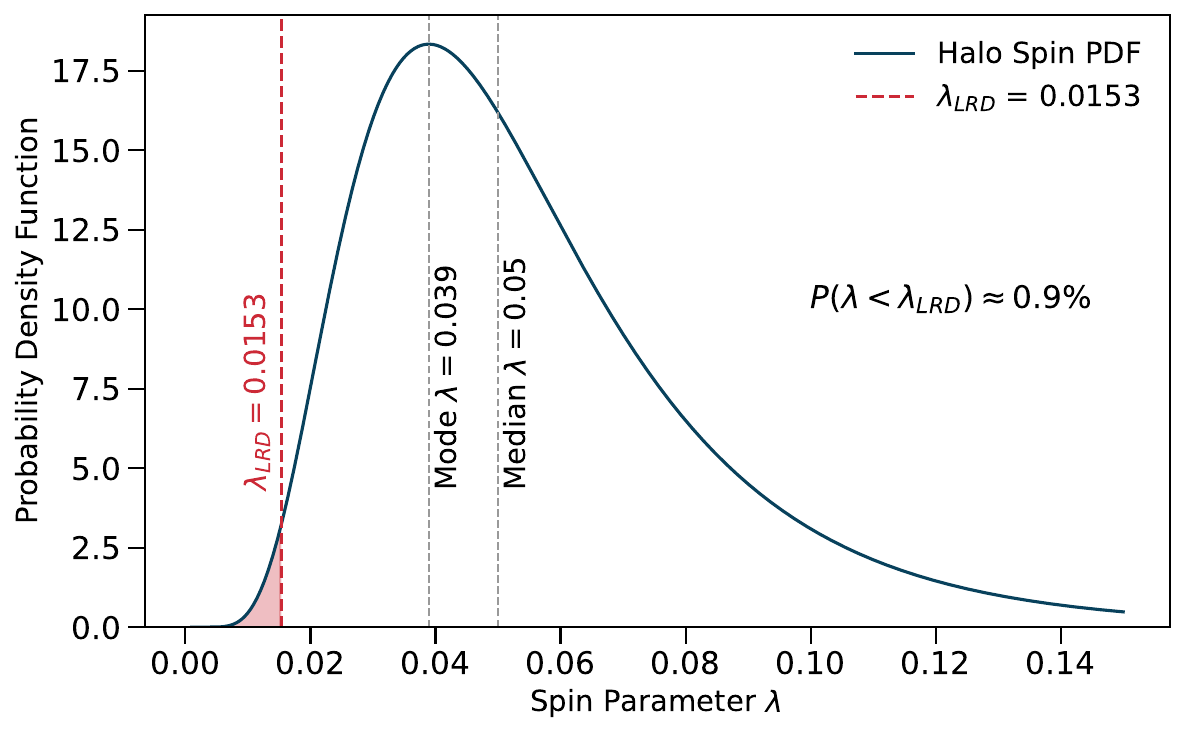}
\caption{Probability density function of the halo spin parameter $\lambda$, modeled as a lognormal distribution \citep{MMW_1998}, with the mode (i.e., the most common spin value) and median indicated for reference. The vertical dashed line marks the critical value $\lambda_{\rm LRD} = 0.0153$, corresponding to the spin threshold required to explain the abundance and compactness of the typical LRD at $z \sim 5$. This value lies in the extreme low-spin tail of the distribution, representing only $\sim 1\%$ of halos. 
\label{fig:spin_pdf}}
\end{figure*}

Small effective radii characterize high-redshift LRDs; \cite{Baggen_2023} found that their average effective radius is $\sim 150$ pc, with a narrow distribution between $\sim 80$ pc and $\sim 300$ pc. Some cases are even more extreme, such as a triply-imaged source at $z = 7.6$ characterized by a de-lensed size upper limit of $R_{\mathrm{eff}} < 35$ pc \citep{Furtak_2023_lensed}. It is currently unclear if the few LRDs identified at $z < 4$ are also characterized by such compactness. Ground-based surveys cannot constrain sizes as well as JWST: for example, at $z \sim 2$, the $0.67$'' PSF of Hyper Suprime-Cam corresponds to $\sim 4$ kpc, rendering all the low-redshift LRD analogs consistent with point sources \citep{Ma_2025}.

For the present purpose, we define a LRD as a compact galaxy with effective radius $R_{\mathrm{eff}} < 300$ pc at $z \sim 5$, which is the median redshift of their bulk distribution \citep{Kocevski_2024}.
We adopt a total galaxy number density \citep{Bouwens_2021} at $M_{\mathrm{UV}}~\sim~-19$ of:
\begin{equation}
\phi_{\mathrm{LBG}} \approx 2.20 \times 10^{-3} \, \mathrm{Mpc}^{-3} \, \mathrm{mag}^{-1} \, ,
\end{equation}
and estimate the number density of LRDs at the same magnitude \citep{Kocevski_2024} to be:
\begin{equation}
\phi_{\mathrm{LRD}} \approx 1.97 \times 10^{-5} \, \mathrm{Mpc}^{-3} \, \mathrm{mag}^{-1} \, .
\end{equation}
This yields a fractional abundance:
\begin{equation}
f_{\mathrm{LRD}}(z=5) = \frac{\phi_{\mathrm{LRD}}}{\phi_{\mathrm{LBG}}} \approx 0.009 \, .
\end{equation}

We now compute the spin threshold $\lambda_{\mathrm{LRD}}$ such that only a fraction $f_{\mathrm{LRD}}$ of halos lie below it in the spin distribution. Using Eq.~(\ref{eq:spin_PDF}), we invert the PDF for $\lambda$, requiring:
\begin{equation}
f_{\mathrm{LRD}} = P(\lambda < \lambda_{\mathrm{LRD}}) = 0.009 \, .
\end{equation}
This yields:
\begin{equation}
\lambda_{\mathrm{LRD}}(z=5) \approx 0.0153.
\end{equation}
Hence, in this framework, the LRDs are hosted in galaxies within the \textit{first percentile of the spin distribution of dark matter halos}, as shown in Fig. \ref{fig:spin_pdf}.

\subsection{Compactness of the LRDs}
\label{subsec:compactness}

We now evaluate whether the same value of $\lambda_{\mathrm{LRD}}$ that explains the abundance of LRDs can also account for their compactness.
Using Eq.~(\ref{eq:size_relation}) to connect spin and effective radius, we evaluate $R_{\rm eff}$ corresponding to $\lambda_{\mathrm{LRD}} = 0.0153$.
For this purpose, we calculate the virial radius $r_{200}(z=5)$ for a halo mass of $M_{\mathrm{halo}} = 10^{11} \Msun$, corresponding to a UV absolute magnitude of $-19$ via abundance matching \citep{Behroozi_2013, Moster_2013}, which is a typical brightness observed in the LRDs \citep{Kocevski_2024, Kokorev_2024_census}. Note that we also used a reference UV absolute magnitude of $-19$ in Sec. \ref{subsec:abundance} to compute the fractional abundance of LRDs as $f_{\mathrm{LRD}}(z=5) = \phi_{\mathrm{LRD}}/ \phi_{\mathrm{LBG}} \approx 0.009$.

With these assumptions, we find:
\begin{equation}
R_{\mathrm{eff}} \approx 260 \, \mathrm{pc} \, .
\end{equation}
Spin values lower than $\lambda_{\mathrm{LRD}}$ lead to effective radii $R_{\mathrm{eff}} < 260$ pc, encompassing most of the observed range of LRD sizes: $80-300$ pc \citep{Baggen_2023}.

\begin{figure*}[ht!]
\plotone{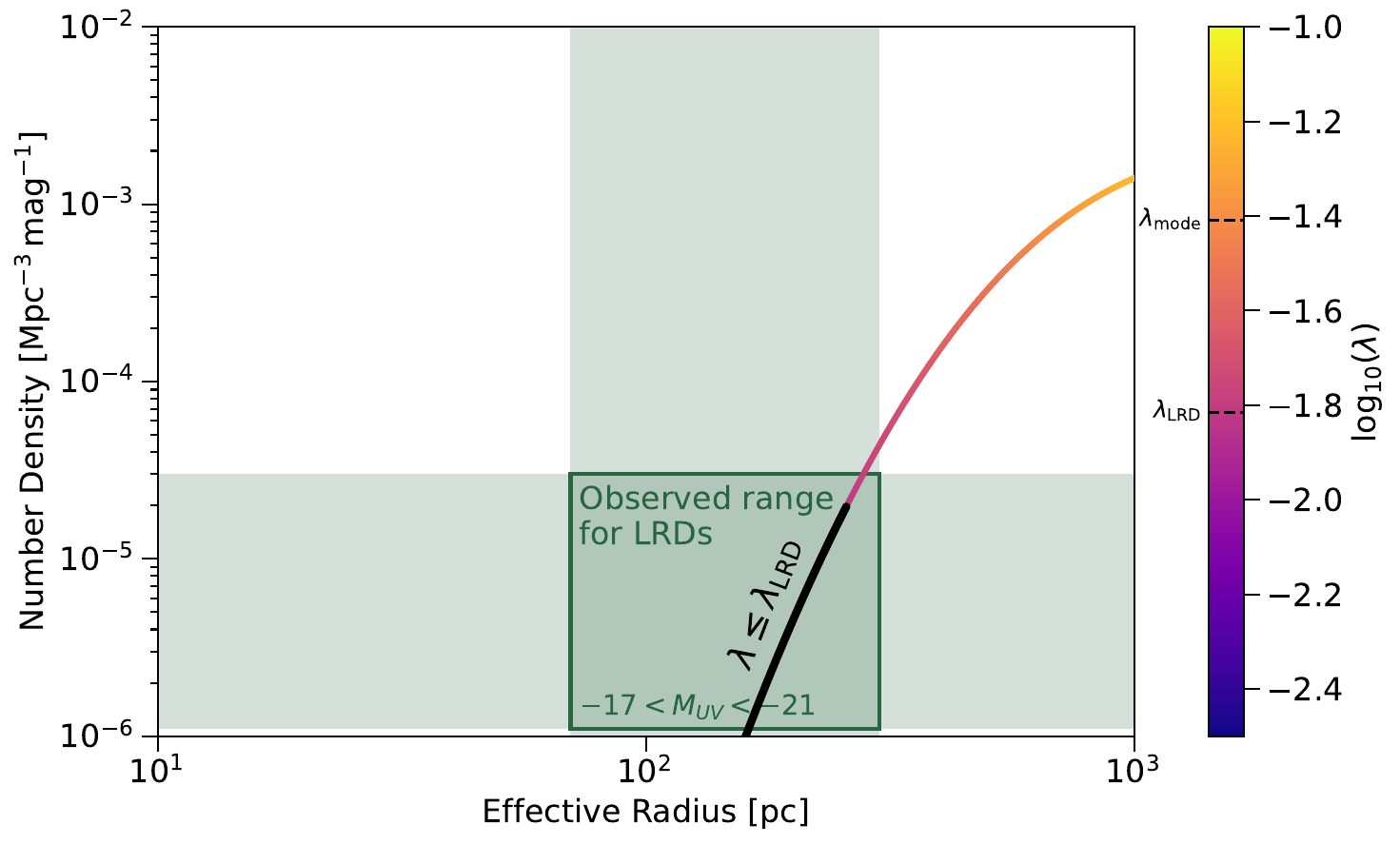}
\caption{Predicted number density of LRDs as a function of effective radius $R_{\rm eff}$ at $z = 5$, for various values of the halo spin parameter $\lambda$, indicated in the color bar. The solid curve shows the theoretical expectation for galaxies with typical UV magnitude $M_{\rm UV} \sim -19$. The shaded horizontal region denotes the observed number density range for LRDs \citep{Kocevski_2024}, while the vertical shaded band marks their observed effective radius range \citep{Baggen_2023}; observational ranges encompass the full brightness range of LRDs, with $-17 <M_{\rm UV} < -21$. The intersection of these shaded regions highlights the regime where low-spin halo galaxies reproduce the abundance and compactness of the observed LRD population.
\label{fig:2D_ranges}}
\end{figure*}

To summarize, the assumption that the typical $z \sim 5$ LRDs are generated from the first percentile of the spin distribution of dark matter halos explains, simultaneously, their observed abundance and compactness, as displayed in Fig. \ref{fig:2D_ranges}. 

Note that because both the effective radius and the abundance are monotonic functions of spin, there is a direct correspondence between them: lower-spin halos produce smaller galaxies, following $R_{\mathrm{eff}} \propto \lambda$, while their abundance scales as $\phi(\lambda) = \phi_{\mathrm{LBG}} \times P(<\lambda)$, where $P(<\lambda)$ is the cumulative distribution function (CDF) of the spin parameter. For $\lambda \ll \bar{\lambda}$, the lognormal CDF behaves approximately like a power law in $\lambda$, so increasingly compact systems become exponentially rarer.

\subsection{Redshift Evolution of the LRDs}
\label{subsec:z_evol}

The redshift distribution of LRDs peaks around $z \sim 5$ and is primarily confined to the range $4 \lesssim z \lesssim 8$ \citep{Kocevski_2024}.: the ``Little Red Dots Era'' of $\sim 1$ Gyr of cosmic time.

However, examples of LRDs are detected at higher (see, e.g., \citealt{Kocevski_2024, Kokorev_2024_census, Taylor_2025}) and lower \citep{Ma_2025, Zhuang_2025} redshifts, albeit in smaller numbers.
It is thus reasonable to conclude that physical and observational constraints hinder the detection of LRDs at redshifts that are lower and higher than the ``LRDs Era''.
This effect is a natural consequence of our theoretical framework, which is centered on the idea that the LRDs originate from low-spin halos.
In fact, the LRDs become \textit{rarely observed}:
\begin{itemize}
    \item at \textit{low redshift} ($z < 4$) because the larger virial radii of halos require extremely low spin values to form compact galaxies, making such configurations increasingly improbable.
    \item at \textit{high redshift} ($z > 8$), because despite being intrinsically more common, their surface brightness drops below JWST's detection threshold due to cosmological dimming.
\end{itemize}
Next, we explore the two regimes separately.

\begin{figure*}[ht!]
\plotone{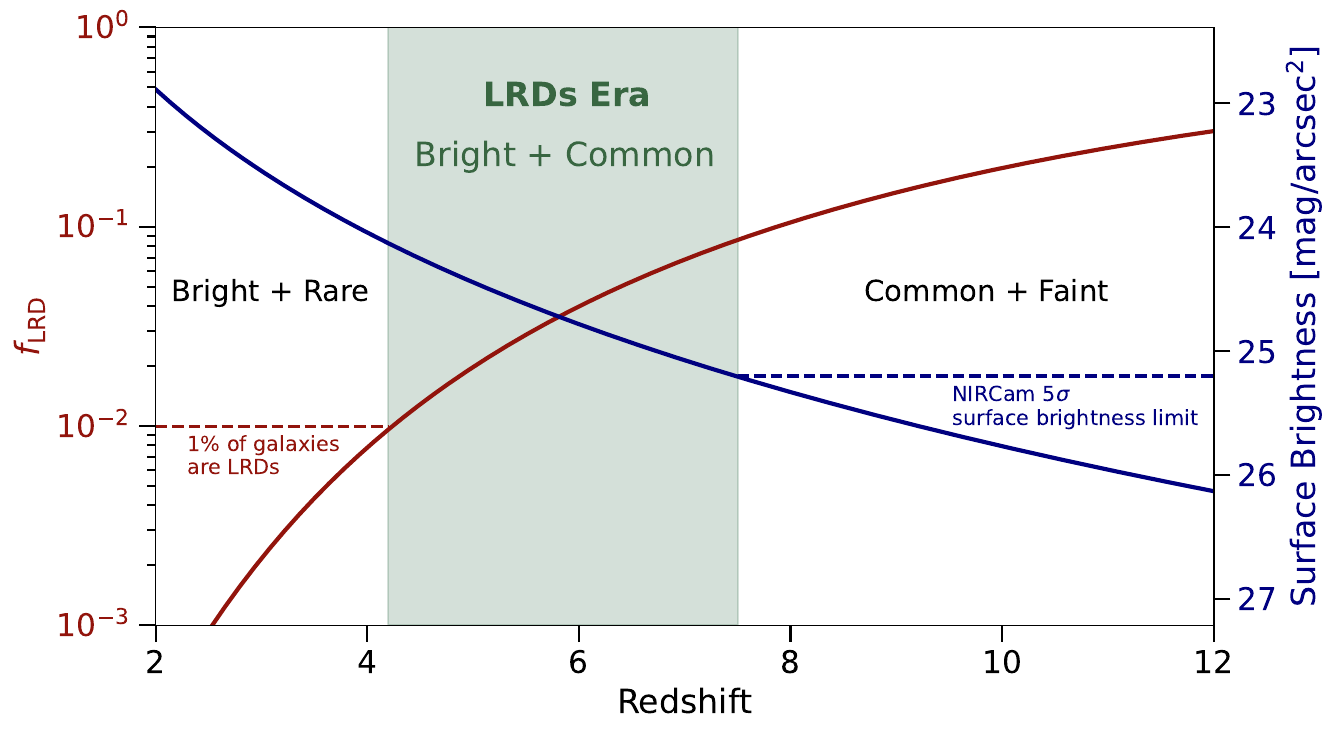}
\caption{This plot illustrates why LRDs are preferentially detected in the redshift range $4 \lesssim z \lesssim 8$.
The red line shows the fraction of halos whose spin parameter falls below the redshift-dependent threshold required to form galaxies more compact than $R_{\mathrm{eff}} = 300$\,pc.
The blue line shows the surface brightness of such a representative compact galaxy.
The dashed blue line marks the assumed JWST/NIRCam 5$\sigma$ surface brightness limit ($\sim$25.2 mag/arcsec$^2$).
The dashed red line shows the compact fraction threshold of $1\%$, which we take as the limit below which LRDs become intrinsically rare.
The red and blue curves serve complementary roles: the former quantifies the intrinsic rarity of compact systems, while the latter estimates their detectability.
The shaded region ($4 \lesssim z \lesssim 8$) highlights the redshift range where LRDs are both common and detectable — the ``LRD Era''.
\label{fig:detectability}}
\end{figure*}

\subsubsection{LRDs at Low Redshifts: Bright + Rare}
We assume that one defining property of the LRDs is their compactness. We then define the compactness threshold at $R_{\mathrm{eff}} = 300 \, \mathrm{pc}$, based on the observed upper size limit of the population \citep{Baggen_2023}. This size threshold corresponds to a redshift-dependent spin threshold, computed using Eq.~(\ref{eq:lambda_LRD}).

As the Universe expands, the average virial radius of dark matter halos increases. Specifically, at fixed halo mass, the virial radius evolves as $r_{200}(z) \propto (1 + z)^{-1}$. Substituting this into Eq.~(\ref{eq:lambda_LRD}), we find that the spin threshold scales as:
\begin{equation}
\lambda_{\mathrm{LRD}}(z) \propto \frac{1}{r_{200}(z)} \propto (1 + z).
\end{equation}
Thus, as redshift decreases, the spin threshold $\lambda_{\mathrm{LRD}}$ becomes smaller.

The consequence of this evolution is that at low redshift, only halos with extremely low spin parameters can produce galaxies that remain below the compactness threshold based on $R_{\mathrm{eff}}$. These low-spin halos become statistically improbable in the tail of the spin distribution: the suppression at low spin values is exponential. 

We quantify this rarity using the cumulative probability:
\begin{equation}
    f_{\mathrm{LRD}}(z) = \int_0^{\lambda_{\mathrm{LRD}}(z)} p(\lambda) \, d\lambda,
\end{equation}
where $p(\lambda)$ is the lognormal spin PDF introduced in Sec.~\ref{sec:theory}.
As shown by the red curve in Fig.~\ref{fig:detectability}, this compact fraction decreases rapidly with decreasing redshift. Note that a larger value of the compactness threshold based on $R_{\mathrm{eff}}$ (say, $500$ pc, or $1$ kpc) would not change the redshift dependence; it would only re-normalize the blue and red curves in Fig. \ref{fig:detectability} upward.

LRDs at low redshift are rarely observed because compact configurations are no longer typical. This characterizes the ``Bright + Rare'' regime: galaxies are bright enough to be detected, but very few halos produce them.

\subsubsection{LRDs at High Redshifts: Common + Faint}
\label{subsubsec:common_plus_faint}

At high redshifts, LRDs are expected to be more common because $f_{\mathrm{LRD}}(z)$ increases with redshift, but they become increasingly difficult to detect. 
Surface brightness, not intrinsic compactness, ultimately determines whether LRDs can be observed in deep JWST imaging.

To estimate the observability of compact galaxies across redshift, we calculate the surface brightness $\mu(z)$ of a representative LRD with spin in the 1st percentile of the halo spin distribution. This choice provides a consistent, physically motivated way to track how the sizes, and thus the surface brightness, of compact systems evolve with redshift. We adopt a practical surface brightness detection threshold of $\mu_{\rm lim} \approx 25.2 \, \mathrm{mag/arcsec}^2$ \citep{Rieke_2023, Finkelstein_2025}, as detailed in Sec.~\ref{sec:theory}.

As shown by the blue curve in Fig.~\ref{fig:detectability}, surface brightness decreases rapidly with redshift due to cosmological dimming and increasing luminosity distance. By $z \gtrsim 8$, even the most compact galaxies in the low-spin tail fall below the detection threshold: $\mu(z) > \mu_{\rm lim}$. This defines the ``Common + Faint'' regime: although LRDs are more common at high redshifts, they are rarely observed because their surface brightness becomes too faint to be detected.

To quantify this observational suppression, we introduce an empirical correction factor $C(z)$ that modulates the predicted LRD abundance based on surface brightness visibility:
\begin{equation}
C(z) = \frac{1}{1 + \exp\left[ \alpha \left( \mu(z) - \mu_{\rm mid} \right) \right]} \, ,
\label{eq:correction}
\end{equation}
where $\mu(z)$ is the surface brightness, $\mu_{\rm mid}$ is the midpoint of the logistic transition (i.e., where $C(z) = 1/2$), and $\alpha$ controls the steepness. We assume $\mu_{\rm mid} = 24.8\, \mathrm{mag/arcsec}^2$ as the typical surface brightness of a compact system in the middle of the ``LRDs Era'' (see Fig. \ref{fig:detectability}). This functional form produces a gradual suppression that begins at $z \sim 5.5$ and becomes severe at $z \gtrsim 8$.

Importantly, this correction factor is partially empirical. While surface brightness dimming is a dominant effect, other observational limitations also contribute to the reduced detectability of LRDs at high redshifts. For example, filter coverage limitations can result in a lack of sufficient photometric bands to apply reliable color selections. This may cause high-$z$ LRDs to be missed or excluded from catalogs during selection procedures. Our correction factor captures the combined impact of these effects and allows us to reproduce the observed turnover in LRD number density at $z \gtrsim 6$.

\subsubsection{The ``LRDs Era'': Bright + Common}

The ``LRD Era'' is defined as the redshift interval in which compact galaxies are both intrinsically common and observationally detectable. Specifically, this corresponds to the range where:
\begin{equation}
f_{\mathrm{LRD}}(z) > 0.01 \quad \text{and} \quad \mu(z) < \mu_{\mathrm{lim}} \, .
\end{equation}

In this regime, the spin threshold $\lambda_{\mathrm{LRD}}(z)$ is high enough that a non-negligible fraction of halos satisfy the compactness condition, and the resulting galaxies are characterized by a surface brightness sufficient to be detected in JWST deep imaging. As shown in Fig.~\ref{fig:detectability}, this window spans approximately $4 \lesssim z \lesssim 8$, where LRDs are both \textit{bright} and \textit{common}. Outside of this range, compact galaxies are either statistically suppressed (at low $z$) or observationally inaccessible (at high $z$), explaining the observed distribution of LRDs in redshift space.

\subsubsection{Redshift Evolution from Observations}

\begin{figure*}[ht!]
\plotone{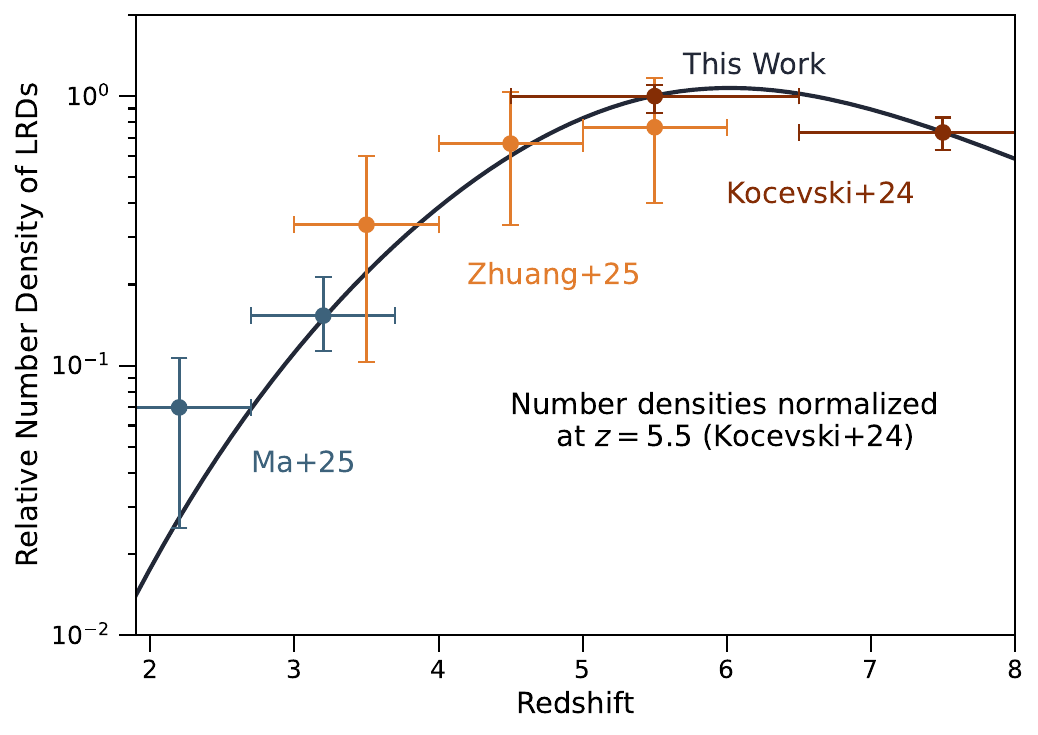}
\caption{Relative number density of LRDs as a function of redshift, normalized to the value at $z=5.5$ from \citet{Kocevski_2024}. 
The solid line shows the theoretical prediction from our model, also normalized at $z=5.5$. 
Colored points show observational estimates from \citet{Kocevski_2024} (brown), \citet{Zhuang_2025} (orange) and \citet{Ma_2025} (blue).
In the regime $z < 5$ when the typical LRD is characterized by detectable surface brightness, the redshift evolution is primarily driven by the redshift evolution of $f_{\rm LRD}$: a lognormal which is exponentially suppressed at low redshifts. At $z \gtrsim 5$, the suppression from cosmological surface brightness dimming becomes essential.
\label{fig:z_evolution}}
\end{figure*}

To evaluate the accuracy of our theoretical predictions, we compare the redshift evolution of the model to observed LRD number densities spanning $2 < z < 8$.
At $z < 5$, the typical LRD is significantly above the estimated $5\sigma$ surface brightness limit; hence, the redshift evolution is primarily driven by the evolution of $f_{\rm LRD}$, shown as a red line in Fig. \ref{fig:detectability}. The correction factor $C(z)$ in Eq. (\ref{eq:correction}) begins to play a crucial role at $z \gtrsim 5$, where cosmological surface brightness dimming becomes essential.

In this comparison, displayed in Fig. \ref{fig:z_evolution}, we normalize all number densities to the observed value at $z = 5.5$ reported by \cite{Kocevski_2024}.
We include two measurements from \citet{Kocevski_2024} at $z=5.5$ and $z=7.5$, three from \citet{Zhuang_2025} at $z=3.5$, $z=4.5$ and $z=5.5$, and two from \citet{Ma_2025} at $z=2.2$ and $z=3.2$. 
The observed trend matches the theoretical prediction very closely: for instance, the LRD number density declines by an order of magnitude from $z=5$ to $z=3$. 

\section{Discussion and Conclusions}
\label{sec:disc_concl}

In this Letter, we have proposed a simple physical framework in which LRDs are naturally interpreted as the descendants of dark matter halos with unusually low spin. We demonstrated that this scenario simultaneously explains the three defining characteristics of the LRD population: their abundance, compactness, and redshift distribution. Our model focuses only on observed, not modeled, properties; it is thus independent of the LRDs being powered primarily by a black hole versus stars, as described in Sec. \ref{sec:introduction}.

Our model relies on a few key assumptions. We define LRDs as galaxies with effective radii smaller than $300$ pc and relate their sizes to the spin parameter of their host halo using the formalism of \cite{MMW_1998}. We adopt a redshift-independent, lognormal spin distribution consistent with cosmological simulations and assess detectability based on the surface brightness limit of deep JWST/NIRCam imaging.

We summarize our main findings below:
\begin{itemize}
    \item The prototypical LRDs at $z = 5$ are explained as galaxies formed in the lowest $\sim 1\%$ of the halo spin distribution. This naturally reproduces both their observed abundance ($\sim 1\%$ of standard galaxies) and their compact sizes ($R_{\mathrm{eff}} < 300$ pc).
    \item The redshift evolution of LRD detections is driven by the interplay between the evolving compact galaxy fraction and cosmological surface brightness dimming. This leads to a well-defined ``LRDs Era'' at $4 \lesssim z \lesssim 8$, during which LRDs are \textit{common and detectable}. At $z < 4$, they are bright enough to be detectable but very rare. At $z > 8$, they are very common but too faint to be detected.
    \item The predicted redshift evolution of the LRD number density naturally matches observations from $z = 8$ to $z = 2$, with an exponentially suppressed lognormal distribution that captures perfectly the steady decline at low redshifts.
\end{itemize}

A range of empirical evidence supports our low-spin model.
First, several studies report that LRDs exhibit excess small-scale clustering (e.g., \citealt{Zhuang_2025}). In our framework, this is a natural outcome: halo spin arises from tidal torques exerted by the surrounding large-scale structure. Regions in which these torques are systematically weak will host halos with low angular momentum, leading to spatially correlated populations of LRDs.
Second, many key spectral features of LRDs imply extremely high stellar and/or gas densities in their cores. For example, broad emission lines are interpreted as arising from extreme stellar densities \citep{Loeb_2024RNAAS, Baggen_2024}, while strong Balmer breaks, absorption features, and the absence of X-ray emission have been attributed to extremely dense gas \citep{Inayoshi_Maiolino_2025, Maiolino_2024_Xray}. These dense core conditions arise naturally in our model since galaxies forming in low-spin halos are not rotationally supported and efficiently funnel mass toward the center.
In such systems, while the dark matter halo is spinning slowly, the baryons cool, lose pressure support, and condense to form a compact, fast-rotating disk. The angular momentum of the disk scales with the square root of its radius, while the disk rotation speed scales inversely with the square root of the disk radius. Since the initial angular momentum per unit mass is low, the disk radius is small, and correspondingly, the gas rotation speed is high.

Low-spin halos may also explain the peculiar red SEDs of the LRDs. In such halos, baryons collapse into a compact central region, leading to high column densities and centrally concentrated star formation. These conditions naturally promote dust formation and obscuration, enhancing emission at longer wavelengths and producing redder colors \citep{Akins_2024, Maiolino_2024_Xray}. Additionally, rapid early star formation in dense cores may contribute to quenching, thereby reinforcing the red signature in their SED (see, e.g., \citealt{Zolotov_2015, Tacchella_2016, Baggen_2024, Guia_2024}) As the system evolves, gas accretion and interactions can build angular momentum, allowing star formation to propagate outward into less obscured, lower-density regions. This outward growth dilutes the central obscuration and adds blue light from younger stellar populations, weakening the red excess. Such a process may explain why the distinctive V-shaped SEDs of the LRDs become less prominent at lower redshifts, consistent with the observed decline of the LRD population beyond the $4 < z < 8$ window.

Although the halo spin distribution is continuous, the resulting galaxy population may appear observationally segmented. In particular, cosmological surface brightness dimming can obscure the extended, low-surface-brightness outskirts of galaxies at high redshifts. A simple calculation for a prototypical LRD with typical central surface brightness shows that, by $z \gtrsim 7$, a substantial fraction of the total light from the galaxy’s disk becomes undetectable, leaving only the bright, compact central region visible.

As a result, high-redshift galaxies with larger, more diffuse morphologies may be entirely missed by current observations if even their central regions fall below the surface brightness detection threshold. In contrast, galaxies forming in low-spin halos naturally concentrate their light into a compact, bright core, making them far more likely to be detected. This visibility bias favors the selection of such systems in JWST deep fields and can lead to an observed population dominated by compact galaxies, even if more extended systems are intrinsically more common.

This selection effect alone can create the appearance of a sharp boundary between compact LRDs and the broader galaxy population. Furthermore, galaxies forming in halos with exceptionally low spin may exhibit unusual structural properties — such as extreme compactness and high central densities — that enhance their contrast with typical high-redshift galaxies. These characteristics can make the LRDs appear as a distinct group, even though their origin lies in the tail of a continuous distribution of halo angular momentum.

In the near future, several avenues offer the potential to test and refine this interpretation. Full spectral fitting of the spatially resolved stellar continuum may reveal whether LRDs indeed lack rotational support, as expected for galaxies formed in low-spin halos. Cosmological simulations such as \textsc{Astrid} \citep{Astrid_BHs, Astrid_galaxy_formation, LaChance_2025} can assess whether compact, red galaxies emerge preferentially in the lowest-spin halo population. Expanded searches for LRDs at $z < 4$ and $z > 8$ will further constrain their redshift evolution, while improved size measurements of low-redshift analogs \citep{Ma_2025} will clarify whether they share the extreme compactness observed at high redshifts.

As JWST continues to push the frontier of discovery in the early Universe, theoretical models will increasingly rise to the challenge, revealing that even the most unexpected populations may have simple, physically motivated origins.

\begin{acknowledgments}
We thank the referee for providing insightful comments on the paper.
F.P. acknowledges fruitful discussions with Josephine Baggen, Dale Kocevski, and Frank van den Bosch. F.P. also acknowledges support from a Clay Fellowship administered by the Smithsonian Astrophysical Observatory. This work was also supported by the Black Hole Initiative at Harvard University, which is funded by grants from the John Templeton Foundation and the Gordon and Betty Moore Foundation. 
\end{acknowledgments}




\bibliography{ms}{}
\bibliographystyle{aasjournal}



\end{document}